\documentclass[aps,prl,reprint,superscriptaddress,showkeys]{revtex4-1}
\usepackage{graphicx}
\usepackage{bm}

\begin{document}

\title{Muon cooling: longitudinal compression}

\author{Yu Bao}
\affiliation{Paul Scherrer Institute, 5232 Villigen--PSI, Switzerland.}

\author{Aldo Antognini}
\email[Corresponding author: ]{aldo@phys.ethz.ch}
\affiliation{Institute for Particle Physics, ETH Zurich, 8093 Zurich, Switzerland.}

\author{Wilhelm Bertl}
\affiliation{Paul Scherrer Institute, 5232 Villigen--PSI, Switzerland.}

\author{Malte Hildebrandt}
\affiliation{Paul Scherrer Institute, 5232 Villigen--PSI, Switzerland.}

\author{Kim Siang Khaw}
\affiliation{Institute for Particle Physics, ETH Zurich, 8093 Zurich, Switzerland.}

\author{Klaus Kirch}
\affiliation{Paul Scherrer Institute, 5232
Villigen--PSI, Switzerland.}
\affiliation{Institute for Particle Physics, ETH Zurich, 8093
Zurich, Switzerland.}

\author{Angela Papa}
\affiliation{Paul Scherrer Institute, 5232 Villigen--PSI, Switzerland.}

\author{Claude Petitjean}
\affiliation{Paul Scherrer Institute, 5232 Villigen--PSI, Switzerland.}

\author{Florian M. Piegsa}
\affiliation{Institute for Particle Physics, ETH Zurich, 8093 Zurich, Switzerland.}

\author{Stefan Ritt}
\affiliation{Paul Scherrer Institute, 5232 Villigen--PSI, Switzerland.}

\author{Kamil Sedlak}
\affiliation{Paul Scherrer Institute, 5232 Villigen--PSI, Switzerland.}

\author{Alexey Stoykov}
\affiliation{Paul Scherrer Institute, 5232 Villigen--PSI, Switzerland.}

\author{David Taqqu}
\affiliation{Paul Scherrer Institute, 5232 Villigen--PSI, Switzerland.}

\begin{abstract}
A 10\,MeV/c $\mu^+$ beam was stopped in helium gas of a few mbar in a
magnetic field of 5\,T. The muon 'swarm' has been efficiently
compressed from a length of 16\,cm down to a few mm along the magnetic
field axis (longitudinal compression) using electrostatic fields. The
simulation reproduces the low energy interactions of slow muons
in helium gas. Phase space compression occurs on the order of
microseconds, compatible with the muon lifetime of 2~$\mu$s. This
paves the way for preparation of a high quality muon beam.

\end{abstract}


\maketitle

Standard muon ($\mu^+$) beams have a relatively high energy and poor
phase space quality.
A new scheme has been proposed making use of stopped $\mu^+$ in a He gas
target~\cite{taqqu}.
Through the stopping process  high intrinsic phase space
compression is achieved.
The remaining challenge is to extract the muons fast enough into vacuum.
This is done by compressing the stopped muon swarm with electric
fields and guiding it into a small extraction hole.
In this paper we report successful demonstration of 
muon swarm compression along the magnetic field lines.

Related schemes have been used in the field of rare isotope
investigations~\cite{bollen08, savard05}.
High energy ion beams are stopped in He gas and manipulated with DC
and RF electric fields to induce drift, radial compression and
extraction from a gas target.
Typical manipulation and extraction times are  5-200~ms.
For muons much faster techniques are vital.

\begin{figure}
\includegraphics[width=0.50\textwidth]{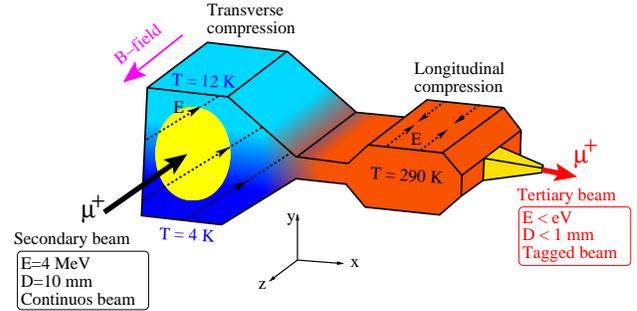}
\caption{\it Schematic of the beam line proposed
  in~\cite{taqqu}. $\mu^+$ traveling in $-z$ direction are stop in 5
  mbar He gas at cryogenic temperatures. First, transverse ($y$
  direction) compression occurs due to density gradient in He gas
  (blue region). Then at room temperature longitudinal ($z$ direction)
  compression takes place (red region). In the yellow region a mixed
  transverse-longitudinal compression takes place before extraction
  into vacuum.}
\label{fig:total-beam-line}
\end{figure}

The new concept for fast compression is based on a position-dependent
muon drift velocity $\vec{v}_D$ in gas.
In a long He gas target placed in a longitudinal high magnetic field, the
stopping muons are first transversely, then longitudinally compressed
and finally extracted through 1~mm$^2$ hole in transverse
direction (see Fig.~\ref{fig:total-beam-line}).
The operation takes place in less than 10~$\mu$s and should lead to a
phase space compression of $10^{10}$.
An intense slow muon beam is obtained with 10~ns time resolution at a
few eV or a micro-beam that can be focused into a beam spot of
10~$\mu$m diameter at 10~keV.
Such a beam can be used for $\mu$SR applications, muonium (Mu)
spectroscopy, muon g-2 experiment, searches for a muon electric dipole
moment (EDM) and Mu-$\mathrm{\overline{Mu}}$ conversion.

The drift velocity of charged particles in gas in the presence of
electric $\vec{E}$ and magnetic $\vec{B}$ fields is~\cite{book}
\begin{equation}
\vec{v}_D=\frac{\mu E}{1+\omega^2\tau_c^2}\Big[\hat{E}+\omega\tau_c \hat{E}\times\hat{B}+\omega^2 \tau_c^2 (\hat{E}\cdot\hat{B})\hat{B} \Big]
\label{eq:v-drift}
\end{equation}
where $\hat{E}$ and $\hat{B}$ are the unit vectors along $\vec{E}$
and $\vec{B}$, $\omega=eB/m$ the cyclotron frequency with $m$ the muon mass,
$\mu$ the muon mobility in the gas and  $\tau_c$ the mean
time between collisions.
For small $\omega \tau_c$ the muon drift is along the electric field
lines.  This is the regime of high gas density where the collision
rate is large.
For large $\omega \tau_c$, $\omega \tau_c\gg 1$,  the third term
dominates and muons follow the magnetic lines.

The beam line we are developing is composed of a
sequence of stages as shown in Fig.~\ref{fig:total-beam-line}.
In the initial {\it transverse} compression stage a $\mu^+$ beam is stopped
in He gas at 5~mbar pressure inside a 5 T longitudinal B-field
$\hat{B}=(0,0,1)$ and a transverse electric field
$\hat{E}=1/\sqrt{2}\,(1,1,0)$.
The gas temperature in this first stage has a vertical gradient from
4~K to 12~K.
At lower densities (top part) $\tau_c$ is large and $\vec{v}_D$ is
dominated by the $\hat{E}\times\hat{B}$ term. Hence the muons drift
diagonally in $-y$ direction and in $+x$ direction.
By contrast, at larger densities (bottom part)
the first term of Eq.~(\ref{eq:v-drift}) dominates and the
resulting drift velocity is along $\hat{E}$.
Therefore, muons originating from the upper (lower) part of the stop
distribution while drifting in $+x$ direction are moving
downwards (upwards) giving rise to a compression in transverse ($y$)
direction.
The 10~mm wide stopping volume in the $y$ direction is thereby reduced
to a swarm of muons moving in $x$ direction with a height of 0.5~mm (in
$y$ direction) and a length (in the $z$ direction) of 50~cm or
more~\cite{taqqu}.

The muon swarm is then entering a second stage at room temperature
(low density) where longitudinal compression (in the direction of
$\hat{B}$ or $z$) takes place. Hence, the muons are directed towards the exit
hole as shown in Fig.~\ref{fig:total-beam-line}.
The third term of Eq.~(\ref{eq:v-drift}) is here dominant because
$\omega \tau_c\gg 1$.
The electric fields are designed such that $E_z$ changes sign in the
center of this region.
Thus, muons drift to the center along the magnetic field lines, which  gives
rise to {\it longitudinal} compression.
A non-vanishing $E_y$ component guarantees that the muon swarm drifts
also along the $+x$ axis because of the $\hat{E}\times\hat{B}$ term.
An additional muon swarm compression in both $y$ and $z$ direction occurs
in a third stage followed by vacuum extraction through a small
orifice.

Longitudinal compression occurs within a relatively large length
of the muon swarm in short time.
This is possible only if at sufficiently high $E_z/N$, where $N$ is
the gas number density, ``runaway'' occurs~\cite{lin}.
As shown by Lin~\cite{lin} for protons, when the kinetic
energy $T$ of the charged particle reaches values greater than 1~eV,
the momentum transfer cross-section $ \sigma_{tr}$ for elastic
scattering (p-He scattering) decreases with increasing $T$ like $1/T$ or faster and
the energy loss due to collisions is not compensating
the energy gain from the electric field and the proton can accelerate
to high energies till other processes come into play.
The onset of ``runaway'' has been demonstrated for protons drifting in
He~\cite{howorka}. Fast longitudinal compression of the $\mu^+$ swarm
within 10-15~cm length requires the $\mu^+$ to be accelerated at
``runaway'' conditions.

\begin{figure}
\includegraphics[width=0.50\textwidth]{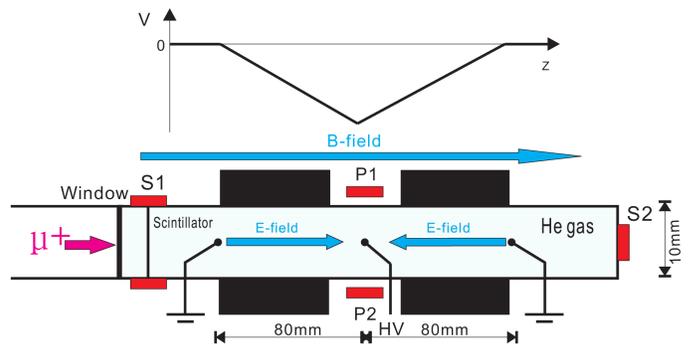}
\caption{\it Setup in the 5T solenoid.  Muons enter the gas target,
  cross the entrance detector $S_1$ and a fraction stop in the He
  gas. The stopped $\mu^+$ are compressed along the B-field using a
  V-shaped electric potential.  $P_1$ and $P_2$ are $e^+$ counters.
  $S_2$ is a scintillator counter for aligning the muon beam. }
\label{fig:setup-2011}
\end{figure}

In this work we present an experimental demonstration of compression
along the magnetic filed lines, {\it i.e.}, longitudinal compression.
A standard $\mu^+$ beam is stopped in a few mbar He gas at
room temperature and 5~T longitudinal magnetic field. The elongated
stopping distribution is compressed into the minimum of a V-shaped
electrostatic potential as shown in Fig.~\ref{fig:setup-2011}.
From the time distribution of the muon-decay positrons ($e^+$) detected with
two positron counters (P$_1$, P$_2$) placed in the vicinity of this
potential minimum it is possible to quantify the muon drift.

The experiment was performed at the $\pi$E1 beamline at the Paul
Scherrer Institute, Switzerland,  tuned to deliver
$\mu^+$ of 10~MeV/c momentum (500~keV energy).
After elimination of the positron contamination with a
electromagnetic separator (Wien filter), the $\mu^+$ beam is focused into a 5~T solenoid containing
a He gas target at a few mbar pressure.
A $\mu^+$ rate of $2\times10^4$~s$^{-1}$ was measured using a
30~$\mu$m thick plastic scintillator  ($S_1$) of 10~mm diameter read out by
four GAPDs (Geiger-mode Avalanche Photo-Diodes).
$S_1$ provides the event start time.

The beam momentum and the scintillator thickness were chosen to give
the highest (few \%) fraction of $\mu^+$ stops in the 160~mm long
``active'' region of the few mbar He target.
The majority of the $\mu^+$ passes through the gas and reaches a second
detector $S_2$ which is also used to align the target along the B-field
lines.

The inner walls of the target are made of printed circuit boards (PCB)
covered by thin metallic strips used to define the V-shaped
electrostatic potential.
To compress the $\mu^+$ to the center ($z=0$) of
the target a negative HV is applied to the central strips.
The other strips are at decreasing absolute potential till $z=\pm
80$~mm where ground potential is applied.
\begin{figure}[t]
\includegraphics[width=0.50\textwidth]{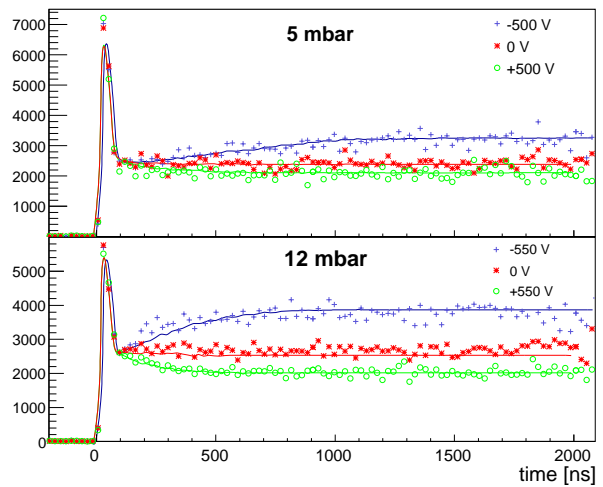}
\caption{\it Measured (continuous lines) and simulated (dashed)
  positron time spectra multiplied with $e^{t/2200}$ for two different
  pressures, where $t$ is the time in ns. The simulations account for
  ``muon chemical capture'' with a rate of $40\cdot 10^6$~s$^{-1}$ and
  a misalignment between target and B-field of 1~mrad (see text).  }
\label{fig:time-spectrum_exp}
\end{figure}

The $S_1$, $S_2$, $P_1$ and $P_2$ signals are recorded with a
DRS4~\cite{ritt} waveform digitizer triggered by
$P_1$ or $P_2$.
Figure~\ref{fig:time-spectrum_exp} shows $e^+$ time
distributions measured in $P_1$ and $P_2$ relative to $S_1$.
To remove the effect related to the $\mu^+$ decay, the histograms have
been scaled with $\mathrm{exp}(t/\tau)$, where $t$ is the time and
$\tau=2.2$~$\mu$s the $\mu^+$ lifetime.
The ``prompt'' peaks originate from the higher energy $\mu^+$ decaying
in flight in the acceptance region (between $\pm 18$~mm in $z$
direction) of $P_1$ and $P_2$.

When a negative HV is applied, the $\mu^+$ are attracted close to $P_1$ and $P_2$.
The resulting $e^+$ time distributions (blue, $+$) show an
increase caused by the muon drift into a region with higher positron
detection acceptance.
On the contrary, applying a positive HV, the $\mu^+$ are pushed
outside the acceptance region resulting in a decrease of
detected positrons (green empty circles).

The curve for large positive HV at delayed time is equivalent to a
background measurement.
Its flat asymptotic behavior reveals that this background is
muon-correlated.
It arises from $\mu^+$ stopping in the target walls, in the beam
collimators, in the beam dump and in the inactive region of the
target.
The background was reduced by only 20\% when emptying the gas cell
(relative to 5 mbar).

A relevant difference between the conditions of this experiment and
those in the longitudinal compression stage of~\cite{taqqu}, is that,
in~\cite{taqqu} the $\mu^+$ enter the longitudinal compression stage
at an energy around 1~eV, while in the experiment presented here, the muons
enter the active volume at keV energies.  Thus, here the slowing-down
process competes and is mixed with the drift induced by the electric field.

To analyze the measured time spectra (which include prompt peak,
background, slowing down and compression) and quantify the
compression, we extended the GEANT4 (version 4.9.3) simulation
package~\cite{geant4} to include $\mu^+$ physics in the 0.1~eV$-$1~keV
energy range.
Below 1~keV energy standard GEANT4 processes have been switched off
and elastic $\mu^+-$He collision, Mu formation and Mu ionization have
been implemented.
We use the data available for protons and we scale them to
muons: velocity scaling is used for the charge-transfer processes and
energy scaling for the elastic collisions~\cite{senba}.
Among these processes the elastic $\mu^+-$He scattering
is the decisive interaction that controls the muon compression.
It has been included in the Monte Carlo (MC) simulation starting from the
proton differential cross sections calculated in~\cite{krstic}.

\begin{figure}
\includegraphics[width=0.50\textwidth]{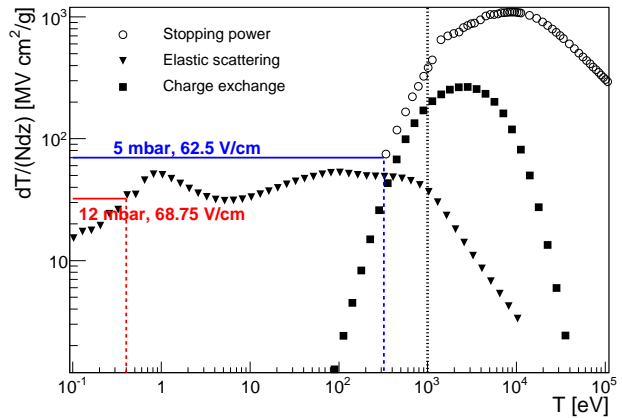}
\caption{\it Energy loss processes of $\mu^+$ at low energy scaled from
  proton data.  The empty circles show the stopping power caused by
  He-ionization, He-excitation and charge exchange processes.  The curve
  with squares accounts for energy losses due to Mu formation and
  ionization.  The curve with triangles represents the kinetic energy
  loss ($dT/(N\,dz)$) caused by elastic scattering per unit length in
  a fixed direction ($z$, electric field direction).  The vertical
  blue and red lines are the resulting equilibrium energies at two
  experimental conditions. }
\label{fig:stopping_power}
\end{figure}

When a $\mu^+$ crosses matter in the keV energy regime it undergoes
charge exchange, that is electron-capture and electron-loss processes.
Both processes have been accounted for separately in our
MC simulation using the velocity scaled cross sections of~\cite{nakai}
with 11~eV energy loss in the electron-capture process, and 13.6~eV in
the electron-loss process and no change in direction for both processes.
The stopping power caused by these processes is given in
Fig.~\ref{fig:stopping_power}, where for simplicity only the net
$\mu^+$ energy losses caused by the full charge exchange cycle is
represented without separating electron-capture and electron-loss.
The $\mu^+$ stopping power, indicated by the circles in
Fig.~\ref{fig:stopping_power}, which results from velocity scaling the
measured proton stopping power~\cite{ziegler}, deviates from the stopping
power given by the charge exchange because it also includes
non-negligible processes as ionization and excitation of the target
helium atom.
Because no cross sections are available for these processes, between
300~eV and 1~keV we have tentatively increased the energy losses of
the charge exchange processes to reproduce the stopping power scaled
from proton data.
This approximation is good enough since simulations show
that the drift of the muon is insensitive to the precise
implementation of these processes, as muons spend only a small fraction of
time in this energy range.

The curve shown with triangles in Fig.~\ref{fig:stopping_power}
illustrates the average kinetic energy ($T$) loss of diffusive motion
per unit length along a given direction ($dz$) when only elastic
scattering takes place.  
This curve has been obtained from the diffusion theory of~\cite{lin}
using the scaled transport cross sections of~\cite{krstic}.
The horizontal red and blue lines represent the kinetic energy gain
per unit displacement in $z$ direction due to an electric
field (in $z$ direction) in the absence of collisions with the He gas.
By equating energy loss caused by collisions, with energy gain due to the
applied electric field, one can estimate the average muon equilibrium
drift energy at the corresponding field and density conditions.
For example, as shown in Fig.~\ref{fig:stopping_power} at $E=68.75$~V/cm
and a pressure of 12~mbar the average kinetic energy is 0.4~eV.
As $E/N$ increases, the average equilibrium energy increases.
If $E/N$ is larger than the local maximum at around 1~eV energy,
muons are accelerated to much higher kinetic energies
as shown by the blue lines.
At this point, the energy gain in the E-field is always larger than the
energy loss due to elastic collisions. 
Hence,  muons are
continuously accelerated (run-away condition) until other processes
(charge exchange and inelastic collisions) limit their speed.
The equilibrium energies computed with our MC simulations based on
differential cross sections reproduce the equilibrium energies
obtained from the diffusion theory in~\cite{lin} and shown in
Fig.~\ref{fig:stopping_power}.

\begin{figure}[tb]
\includegraphics[width=0.50\textwidth]{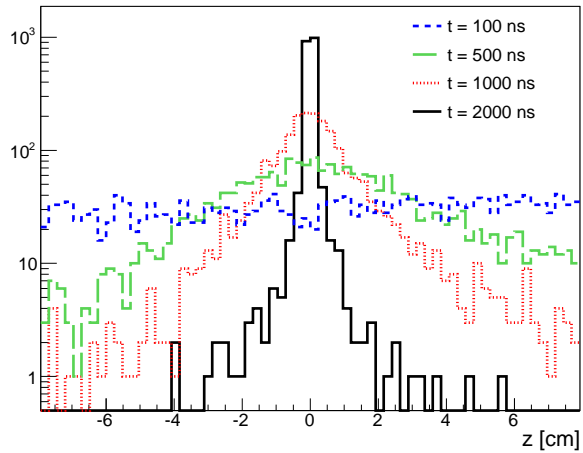}
\caption{\it Muon distribution at various times during the compression
  for 5 mbar and HV=-550 V. }
\label{fig:muon_swarm_vs_time}
\end{figure}

A simulated time evolution of the muon spatial distribution during compression in
the target is given in Fig.~\ref{fig:muon_swarm_vs_time}.
Starting from an approximately flat stop distribution ($t=100$~ns),
the muons are compressed in the center.
Some broadening of the peak at intermediate times is caused by muons
which have already reached the center but are still oscillating while
thermalizing around the minimum of the V-shape potential.
Figure~\ref{fig:muon_swarm_vs_time} shows that it takes less than
2~$\mu$s to compress the stopping muons in our target within $\pm
2$~mm.

\begin{figure}[t]
\includegraphics[width=0.50\textwidth]{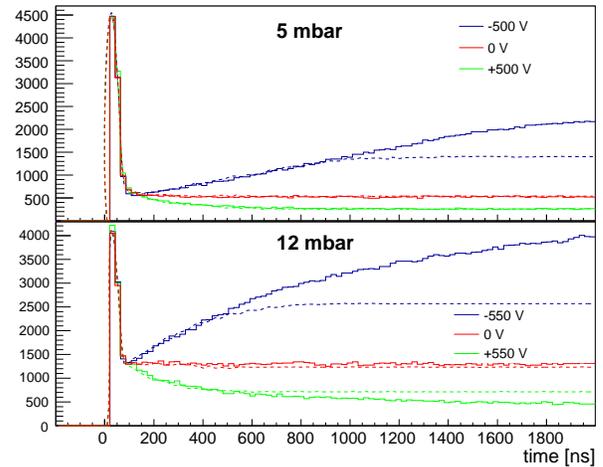}
\caption{\it Simulated $e^+$ time spectra without (continuous
  lines) and with (dashed lines) muon ``chemical capture'' to
  impurities with $R = 40\cdot 10^{6}$~s$^{-1}$ and 10~eV cut-off. 
}
\label{fig:geant4-compression-time-distribution}
\end{figure}

The positron time spectra of $P_1$ and $P_2$ simulated with our MC are
shown in Fig.~\ref{fig:geant4-compression-time-distribution}.
The 12 mbar with HV$=-550$~V (continuous) curve has a fast and a slow component.
The slow component which becomes visible at times later than 700~ns arises
from slow muons drifting at the equilibrium energy of 0.8~eV.
The fast component results from compressing muons  while having several
hundreds eV energy, thus, still in the process of slowing down to the
equilibrium energy.
For 5~mbar no fast and slow component can be distinguished since the
equilibrium drift energy is higher (around 200~eV) and the muon
deceleration process from the 100~keV energy regime is slower. Hence, most muons
reach the center of the target before they slow down.

The measured time spectra of Fig.~\ref{fig:time-spectrum_exp} do not
show any hint of a slow component.
At 12 mbar the measured compression is completed at about 500~ns.
According to simulations, at this time the $\mu^+$ energy ranges from
1~eV to 10~eV.
The probable explanation for the absence of the slow component is that at these
energies the muon interacts with impurities, organic molecules or
water, to form muonic ions or to replace a proton of these molecules
("chemical capture").
The origin of this interaction is the polarization potential 
between the $\mu^+$ and the impurities.
It attracts the muon from a far distance into the atom with a cross
section varying as 1/velocity at low energies.
This results in an interaction rate independent on the energy.
Above a certain energy the muon may interact with  the molecule without
``chemical capture''.
We therefore introduced into the MC a constant loss rate $R$ in the low energy
range up to a cut-off energy which will be taken to be 10~eV.
Large desorption rates have been measured in our target (PCB glued with
araldite) giving rise to  impurities of about
$10^{14}$ molecules/cm$^3$ corresponding  to few atomic
monolayers when integrated over the active target length.
It is known that few
monolayers of impurities on the Ar-moderator surface of the LEM beam
line~\cite{prokscha} are sufficient to reduce by one order of magnitude the
number of $\mu^+$ with energy between 0 to 30~eV emerging from the
surface~\cite{morenzoni}.
The dotted curves in
Fig.~\ref{fig:geant4-compression-time-distribution} show the
effect of this ``chemical capture'' assuming a rate of $R=40\cdot
10^{6}$~s$^{-1}$ for energies $<10$~eV.
These values where chosen to fit the measured data of
Fig.~\ref{fig:time-spectrum_exp}.

The measured time distributions show a larger flat (muon correlated)
background than the simulations.
This increased background can be explained by a small (within
experimental uncertainties) misalignment between the magnetic field
lines and the target, causing an increased number of $\mu^+$ stopping
in the target walls in front of the positron detectors.
Adjusting the simulations with such misalignment and including
``chemical capture'', yields complete agreement with the measured
spectra as shown in Fig.~\ref{fig:time-spectrum_exp}.
Variation of the elastic $\mu^+-$He cross section by a factor of 2
leads to simulated time spectra which disagree strongly with
the measurements in terms of compression speeds.

In conclusion, for the first time we have demonstrated the compression
of stopping muons along the magnetic field direction.
This demonstration relies on the agreement between the experimental
results and simulations based on GEANT4 extended to account for low
energy processes.
For the first time run-away conditions for muons~\cite{lin} have been
experimentally exploited.
Applying our MC to muons starting the longitudinal compression process
at low energy as in~\cite{taqqu}, we predict a longitudinal
compression process to take place with almost no losses in walls or
muonium formation in less than 2 $\mu$s (for a swarm length of
160~mm).

\begin{acknowledgments}
This work was performed at the proton accelerator of the Paul Scherrer
Institute, Villigen, Switzerland. We are grateful to machine and
support groups whose outstanding effort has made this experiment
feasible.  The work was supported by SNF grant 200020\_146902.
We acknowledge discussions and help of F. Kottmann, R.~Scheuermann,
T.~Prokscha, D.~Reggiani, K.~Deiters, T.~Rauber, O.~Kiselev, Y.~Lee,
M.~Diepold, Z.~Hochmann and F. Barchetti.
\end{acknowledgments}

\end{document}